\documentclass[aps, prd, superscriptaddress,preprintnumbers,nofootinbib]{revtex4-1}

\usepackage[english]{babel}
\usepackage[T1]{fontenc}
\usepackage[utf8]{inputenc}
\usepackage{xcolor}
\usepackage{verbatim}
\usepackage{empheq}
\usepackage{amssymb}
\usepackage{amsmath}
\usepackage{amsfonts}
\usepackage{float}
\usepackage{nicefrac}
\usepackage{appendix}
\usepackage{cancel}
\usepackage{soul}

\usepackage[colorlinks]{hyperref}
\usepackage[noabbrev]{cleveref}
\hypersetup{citecolor=blue,linkcolor=blue,urlcolor=blue}

\usepackage{dcolumn}
\usepackage{graphicx}
\usepackage{etoolbox}
\usepackage{booktabs}
\usepackage{babel}

\usepackage{tikz} 
\usetikzlibrary{3d} 
\usetikzlibrary{calc} 
\usetikzlibrary{decorations.pathreplacing} 
\usepackage{pgfplots}

\newcommand\pgfmathsinandcos[3]{
	\pgfmathsetmacro#1{sin(#3)}
	\pgfmathsetmacro#2{cos(#3)}
}
\newcommand\LongitudePlane[3][current plane]{
	\pgfmathsinandcos\sinEl\cosEl{#2} 
	\pgfmathsinandcos\sint\cost{#3} 
	\tikzset{#1/.style={cm={\cost,\sint*\sinEl,0,\cosEl,(0,0)}}}
}
\newcommand\LatitudePlane[3][current plane]{
	\pgfmathsinandcos\sinEl\cosEl{#2} 
	\pgfmathsinandcos\sint\cost{#3} 
	\pgfmathsetmacro\yshift{\cosEl*\sint}
	\tikzset{#1/.style={cm={\cost,0,0,\cost*\sinEl,(0,\yshift)}}} 
}
\newcommand\DrawLongitudeCircle[2][1]{
	\LongitudePlane{\angEl}{#2}
	\tikzset{current plane/.prefix style={scale=#1}}  
	\pgfmathsetmacro\angVis{atan(sin(#2)*cos(\angEl)/sin(\angEl))} %
	\draw[current plane,thin,black] (\angVis:1) arc (\angVis:\angVis+180:1);
	\draw[current plane,thin,dashed] (\angVis-180:1) arc (\angVis-180:\angVis:1);
}
\newcommand\DrawLatitudeCircle[2][1]{
	\LatitudePlane{\angEl}{#2}
	\tikzset{current plane/.prefix style={scale=#1}}
	\pgfmathsetmacro\sinVis{sin(#2)/cos(#2)*sin(\angEl)/cos(\angEl)}
	\pgfmathsetmacro\angVis{asin(min(1,max(\sinVis,-1)))}
	\draw[current plane,thin,black] (\angVis:1) arc (\angVis:-\angVis-180:1);
	\draw[current plane,thin,dashed] (180-\angVis:1) arc (180-\angVis:\angVis:1);
}
\tikzset{
	>=latex,
	inner sep=0pt,
	outer sep=2pt,
	mark coordinate/.style={
		inner sep=0pt,
		outer sep=0pt,
		minimum size=3pt,
		fill=black,circle}
}
\pgfplotsset{
	compat=newest,
	colormap={mycolormap}{color=(lightgray) color=(white) color=(lightgray)}
}

\usepackage{xcolor}

\newcommand{\ra}{\rangle}

\newcommand{\sumint}{\;\; \mathclap{\displaystyle\int}\mathclap{\textstyle\sum} \;\;\;}

\newcommand{\orderOf}[1]{\mathcal{O}( #1 )}
\newcommand{\disc}{\text{Disc}}

\newcommand{\wilsonCoeff}[1]{\mathcal{C}_{#1}}

\newcommand{\im}{i}

\newcommand{\ie}{\emph{i.e.}\;}
\newcommand{\ig}{\emph{i.g.}\;}
\newcommand{\cf}{\emph{c.f.}\;}
\newcommand{\amp}{\mathcal{A}}

\newcommand{\bea}{\begin{eqnarray}}
\newcommand{\eea}{\end{eqnarray}}

\def\beq{\begin{equation}}
\def\eeq{\end{equation}}

\newcommand{\nc}{\newcommand}

\newcommand{\cA}{\mathcal{A}}
\newcommand{\cB}{\mathcal{B}}
\newcommand{\cC}{\mathcal{C}}
\newcommand{\cD}{\mathcal{D}}
\newcommand{\cE}{\mathcal{E}}
\newcommand{\cF}{\mathcal{F}}

\newcommand{\cL}{\mathcal{L}}
\newcommand{\cM}{\mathcal{M}}

\newcommand{\eg}{\emph{e.g.}\;}

\nc{\vp}{\phi}
\nc{\tvp}{\widetilde{\phi}}
\nc{\vpj }{\mbox{${\vp^\dag i\,\raisebox{2mm}{\boldmath ${}^\leftrightarrow$}\hspace{-4mm} D_\mu\,\vp}$}}
\nc{\vpjt}{\mbox{${\vp^\dag i\,\raisebox{2mm}{\boldmath ${}^\leftrightarrow$}\hspace{-4mm} D_\mu^{\,a}\,\vp}$}}

\definecolor{Orange}{cmyk}{0,0.61,0.87,0}
\definecolor{JungleGreen}{cmyk}{0.99,0,0.52,0}
\definecolor{OliveGreen}{cmyk}{0.64,0,0.95,0.40}
\definecolor{Brown}{cmyk}{0,0.81,1,0.60}
\definecolor{RoyalBlue}{cmyk}{0.71,0.53,0,0.12}

\newcommand{\lb}[1]{\textcolor{Orange}{LB: #1}}

\newcommand{\pac}[1]{\textcolor{Brown}{}}
\newcommand{\pa}[1]{\textcolor{Brown}{}}

\begin{document} 

\begin{center}
\hfill  TUM-HEP-1558/25

\vspace{2.0cm}

\global\long\def\order#1{\mathcal{O}{\left(#1\right)}}
\global\long\def\d{\mathrm{d}}
\global\long\def\P{P}
\global\long\def\amp{{\mathcal M}}

\title{Renormalising the Field-Space Geometry}

\author{Patrick Aigner}
\email[Electronic address: ]{patrickaigner@gmx.de}
\affiliation{Technische Universität München, Physik-Department, James-Franck-Strasse 1, 85748 Garching,
Germany}

\author{Luigi Bellafronte}
\email[Electronic address: ]{lbellafronte@fsu.edu}
\affiliation{Physics Department, Florida State University, Tallahassee, FL 32306-4350, USA}

\author{Emanuele Gendy}
\email[Electronic address: ]{emanuele.gendy@tum.de}
\affiliation{Technische Universität München, Physik-Department, James-Franck-Strasse 1, 85748 Garching,
Germany}

\author{Dominik Haslehner}
\email[Electronic address: ]{dominik.haslehner@tum.de}
\affiliation{Technische Universität München, Physik-Department, James-Franck-Strasse 1, 85748 Garching,
Germany}
\affiliation{Max Planck Institute for Physics,  Boltzmannstrasse 8, 85748 Garching, Germany}
\author{Andreas Weiler}
\email[Electronic address: ]{andreas.weiler@tum.de}
\affiliation{Technische Universität München, Physik-Department, James-Franck-Strasse 1, 85748 Garching,
Germany}

\begin{abstract}

We present a systematic study of one-loop quantum corrections in scalar effective field theories from a geometric viewpoint, emphasizing the role of field-space curvature and its renormalisation. By treating the scalar fields as coordinates on a Riemannian manifold, we exploit field redefinition invariance to maintain manifest coordinate independence of physical observables. Focusing on the non-linear sigma model (NLSM) and \(\phi^4\) theory, we demonstrate how loop corrections induce momentum- and scale-dependent shifts in the curvature of the field-space manifold. These corrections can be elegantly captured through the recently proposed geometry-kinematics duality, which generalizes the colour-kinematics duality in gauge theories to curved field-space backgrounds. Our results highlight a universal structure emerging in the contractions of Riemann tensors that contribute to renormalisation of the field-space curvature. In particular, we find explicit expressions and a universal structure for the running curvature and Ricci scalar in simple models, illustrating how quantum effects reshape the underlying geometry. This geometric formulation unifies a broad class of scalar EFTs, providing insight into the interplay of curvature, scattering amplitudes, and renormalisation.
\end{abstract}

\maketitle
\end{center}
\newpage

\section{Introduction}

Quantum field theories (QFTs) and effective field theories (EFTs) formulated in terms of a Lagrangian exhibit an inherent invariance under field redefinitions at the level of the action, ensuring that different parametrisations of the fields lead to equivalent physical predictions~\cite{Chisholm:1961tha,Kamefuchi:1961sb,Politzer:1980me,Arzt:1993gz,cohen2024fieldredefinitionsnonlocal}. This invariance can obscure the underlying simplicity of the theory, as physical observables derived from the S-matrix must remain unchanged under such field redefinitions. Mathematically, these field redefinitions can be interpreted as coordinate changes on the field-space manifold, where the fields themselves serve as coordinates. This connection between field redefinitions and the geometry of the field-space manifold is well established and has become a standard technique for theories of scalars~\cite{Meetz:1969as,Honerkamp:1971xtx,Honerkamp:1971sh,Ecker:1971xko,Volkov:1973vd,Tataru:1975ys}. The geometrical picture of field redefinitions has seen recent interest as it can simplify calculations~\cite{Alonso:2015fsp,Alonso:2016oah,Alonso:2017tdy,Helset:2018fgq,Helset:2020yio,Cohen:2020xca,Alonso:2022ffe,Alonso:2015fsp,Alonso:2016oah,Helset:2022pde,Jenkins:2013zja,Jenkins:2013wua,Alonso:2013hga,Chala:2021pll,DasBakshi:2022mwk,Helset:2022pde,DasBakshi:2023htx,Helset:2022pde,Assi:2023zid,Polyakov:2010pt,Buchalla:2019wsc,Helset:2020yio,Helset:2022pde,Cohen:2022uuw,Craig:2023hhp,Alminawi:2023qtf,Alonso:2021rac,Alonso:2022ffe,jenkins2023looprenormalizationscalartheories},  and it has been used to compute renormalisation group equations that depend on the curvature of field-space in SMEFT. Since physical observables and the S-matrix, encode fundamental properties of the theory, they must be independent of any particular choice of coordinates on the field-space manifold $\cM$ and instead depend only on coordinate-invariant properties of the manifold, such as curvature, rather than any specific choice of field parametrisation~\cite{Volkov:1973vd}. By adopting a geometric formulation of physical observables, one can avoid reliance on particular field choices, ensuring that all descriptions remain manifestly coordinate-independent and gaining a deeper understanding of the intrinsic structure of the theory, such as geometric soft theorems~\cite{Cheung:2021yog,derda2024softscalarseffectivefield}. This geometric picture of EFTs has seen recent interest~\cite{Alonso:2015fsp,Alonso:2016oah,Alonso:2017tdy,Finn_2020,Helset:2018fgq,Helset:2020yio,alonso2023primerhiggseffectivefield,Hays:2020scx,Cohen:2020xca,Corbett:2021eux,Corbett:2021cil,Cohen:2021ucp,Martin:2023fad,Gattus:2023gep,Cohen:2021ucp,Li:2024ciy} and has been well studied and understood for field redefinitions that do not include derivatives of the fields. The geometric framework has been generalised in several directions to include particles with spin and field redefinitions with derivatives~\cite{Finn:2020nvn,Cheung:2022vnd,Cohen:2022uuw,Helset:2022tlf,Craig:2023wni,Cheung:2022vnd,Craig:2023hhp,Assi:2023zid,Alminawi:2023qtf,cohen2023amplitudesfieldredefinitions,Cohen:2024bml}.

In this work we want to restrict ourselves to the discussion of the Non-Linear Sigma Model (NLSM) Lagrangian and study how quantum corrections change the curvature of the field-space manifold. The field-space metric $g_{ab} (\phi)$, as other bare parameters in the Lagrangian, is subject to quantum corrections resulting from loops and in particular we study the renormalisation-group flow of the curvature tensor and see how the field-space geometry changes at different energy scales. This momentum-dependent curvature tensor can be interpreted in the context of the recently discovered geometry-kinematics duality~\cite{Cheung:2022vnd}. Understanding the geometry of field-space in NLSMs provides insights into quantum field theories, bridging concepts from differential geometry and physics.

We begin with a concise review in Section \ref{sec_primer}. In Section \ref{sec_oneLoop}, we detail the one-loop quantum effects. 
 
Finally, in Section \ref{sec_curv} we extend the geometry-kinematics duality to include a geometrised local counter-term and discuss the running of the geometric quantities. 

In Appendix \ref{app_OnShell}, we provide a cross-check to the 1-loop amplitude via on-shell methods and in Appendices \ref{ssec_contractions} and \ref{app_2Loop} we argue for a unique 1-loop and 2-loop structure.

\section{Primer}
\label{sec_primer}

In quantum field theory, the notion of a \emph{field-space manifold} arises naturally in theories with multiple interacting scalar fields, such as the Nonlinear Sigma Model (NLSM). Here, the possible field values at each point in spacetime are treated as coordinates on a differentiable manifold, endowing the theory with rich geometric features. The NLSM Lagrangian typically takes the form

\begin{equation}
	\cL = \frac{1}{2} \, g_{ab} (\phi) \, \partial \phi^a \partial \phi ^b\, ,
\end{equation}

where $g_{ab} (\phi)$ is the metric of the field-space manifold $\cM$, inducing a geometric structure on the field-space and interactions between the scalars. In contrast to a free scalar field theory, whose field-space metric is flat and euclidean, the non-trivial geometry of $g_{ab} (\phi)$ induces interactions among the scalars.

As always, we are free to redefine our fields, in the NLSM this is very well captured in the fact that $1^\text{st}$ order field derivatives transform like elements of the tangent space and thus allow the form of the Lagrangian to be field redefinition invariant\footnote{As long as these transformations do not include field derivatives, the transformation properties match those of elements of the tangent space of a Riemannian manifold.}. For transformations $\phi\rightarrow \tilde{\phi}$ we observe that indeed the partial derivatives of $\phi$ with respect to the spacetime coordinate transform as $\partial_\mu \tilde{\phi}^a = \frac{\partial \tilde{\phi^a}}{\partial \phi^b} \,  {\partial_\mu \phi^b}$, which is the transformation behaviour of an element of the tangent space of a Riemannian manifold under coordinate change $\phi\rightarrow \tilde{\phi}$. One observes that the change in $g_{ab}(\phi)$ is given by $\tilde{g}_{ab}(\phi) =\frac{\partial\phi^c}{\partial\tilde{\phi}^a}\frac{\partial\phi^d}{\partial\tilde{\phi}^b} \, g_{cd} (\phi)$, mirroring the transformation properties of a $(0,2)$ - tensor on a Riemannian manifold. Since this tensor maps two elements of the tangent space onto a real number, is symmetric, and exhibits the correct transformation properties, we can think of it as a metric on a Riemannian manifold, the \emph{field-space manifold}. We can now expand the metric about the vacuum configuration of the scalar fields and think of the fields $\phi^a$ as the coordinates on the field-space manifold and choose the vacuum $\vec{\phi}_0$ to be situated at the origin. With this, we can express the metric as a Taylor series about the origin (\ie the vacuum configuration of the fields)

\begin{align}
\label{eq_metricTaylor1}
	g_{ab} ({\phi}) = g_{ab}(0) + \partial_c\, g_{ab}(0) \,\phi^c + \frac{1}{2} \partial_c \partial_d\, g_{ab}(0) \,\phi^c\phi^d + \ldots \, . 
\end{align}

With this Taylor expansion, we see the interactions arising as higher dimensional operators in the NLSM. Since we have a Riemannian structure on the field-space manifold, we know that coordinate transformations will not affect the physics, so we can choose any coordinate system to simplify the Langrangian. Since the field-space is equipped with a metric, we can define geometric quantities such as the Levi-Civita connection, curvature tensors, and geodesics and we can then choose Riemann Normal Coordinates (RNC) to simplify the Taylor expansion about the vacuum. Choosing RNC amounts to demanding $g_{ab}(0)= \delta_{ab}$, $\nabla_a |_0 = \partial_a$, and $\partial_c \, g_{ab}= 0$. In RNC the Christoffel symbols vanish at the origin, \ie $\Gamma|_{(0)}=0$ and the second derivative of the metric is related to the Riemann curvature tensor via 

\begin{equation}
\label{eq_metricRiemannRel}
	\partial_c \partial_d \, g_{ab}(0) = \frac{1}{3}\,\big(R_{acdb} + R_{adcb} \big) _{(0)}\, .
\end{equation}

In RNC, the Riemann curvature tensor is given by

\begin{align}
 R_{abcd}(0)=  \frac{1}{2} ( g_{ad,bc}-g_{ac,bd}  + g_{bc,ad} - g_{bd,ac})_{(0)} \, ,
\label{rimt}
\end{align} 

where we have introduced the notation
\begin{equation}
\partial_c \partial_d \, g_{ab}(0) = g_{ab,cd}(0) \, .
\end{equation}

\noindent The NLSM Lagrangian then simplifies to

\begin{align}
\begin{split}
\label{eq_LaNLSMinRNC1}
	\cL = \frac{1}{2} \Bigg(\delta_{ab} &+ \frac{1}{3} R_{acdb}(0)\, \phi^c\phi^d +\frac{1}{6} \nabla_e\, R_{acdb}(0)\, \phi^c\phi^d\phi^e \\ &+\bigg(\frac{2}{45} \bigl( R_{acdf}{R^f}_{egb}\bigl)+ \frac{1}{20} \nabla_e\nabla_g\,  R_{acdb}\bigg)_{(0)}\,\phi^c\phi^d\phi^e\phi^g + \ldots\Bigg) \partial \phi^a \partial \phi^b \, ,
\end{split}
\end{align} 

\noindent where $R_{abcd}$ is the Riemann curvature tensor and $\nabla_a$ the covariant derivative on the field-space manifold. Here we see that the non-trivial field-space geometry leads to interactions of the fields arising from the curvature tensor and covariant derivatives thereof. On-shell amplitudes are now fully determined by the field-space geometry and the kinematics of the particles, \eg the on-shell $4$-point and $5$-point amplitudes~\cite{Cheung:2021yog} are given by

\begin{align*}
    {\cA^4}_{abcd} &= R_{abcd} \, t + R_{acbd}\,  s, \\
    {\cA^5}_{abcde} &= \nabla_c R_{adbe} \, s_{45} + \nabla_d R_{acbe} \, s_{35} + \nabla_d R_{abce} \, s_{25} \notag \\
    &\quad + \nabla_e R_{acbd}\,  s_{34} + \nabla_e R_{abcd}\, (s_{24} + s_{45})\, .
\end{align*}

Because the metric tensor and elements of the tangent space transform consistently under coordinate changes on the field-space manifold, the expressions for the amplitudes remain valid in any coordinate system we use to describe the field-space manifold. The field-space redefinition invariance of a theory is a consequence of the diffeomorphism invariance of maps on a Riemannian manifold. The amplitudes do not depend on the arbitrary coordinate system we choose to describe the field-space manifold (or write down the Lagrangian in), but they are fully determined by the coordinate-independent geometry of the field-space manifold, expressed here via the curvature tensor and its covariant derivatives.

Building on colour-kinematics duality, in which the structure constants $f^{abc}$ of the gauge group are mapped to momentum dependent functions that still obey Jacobi identities (and hence describe gravity as a gauge theory where colour has been replaced by kinematics) the authors of~\cite{Cheung:2022vnd} proposed a geometry-kinematics duality. Since the NLSM is governed by the field-space metric and its derivatives, promoting the geometrical quantities to momentum-dependent objects, broadens the framework to describe arbitrary theories of massless bosons. The duality is defined on the level of the NLSM action, where the momentum-dependent metric coefficients are determined by

\begin{align}
\begin{split}
\label{eq_actionDuality}
	S = -\frac{1}{2} \int\limits_{p_1, p_2} (p_1 \cdot p_2) \phi_{a}(p_1) \phi_{b}(p_2) 
		\Big[ \delta_{ab} \delta(p_{12}) &+ \int\limits_{p_3} g_{ab|c}(p_1, p_2 | p_3) \delta(p_{123}) \phi_c(p_3) \\
		&+ \frac{1}{2} \int\limits_{p_3, p_4} g_{ab|cd}(p_1, p_2 | p_3, p_4) \phi(p_3) \phi(p_4) \delta(p_{1234}) + \cdots \Big],
\end{split}
\end{align}
with the duality replacement
\begin{align}
\begin{split}
\label{eq_geoKinDualityMap}
	g_{ab} &\to g_{ab}\big(p_1,p_2\big) \delta(p_{1}+p_2) \, ,  \\
	g_{ab,c} &\to g_{ab|c}\big(p_1,p_2 | p_3\big) \delta(p_{1}+p_2+p_3) \, ,  \\
	g_{ab,cd} &\to g_{ab|cd}\big(p_1,p_2 | p_3, p_4\big) \delta(p_{1}+p_2+p_3+p_4) \, .
\end{split}
\end{align}

\noindent With these replacements, the colour Riemann tensor is mapped to a colour and kinematics-dependent version given by

\begin{align}
\begin{split}
\label{eq_genKinCurvTensor}
	R^\text{colour}_{abcd} \to &R_{abcd}({p_1},{p_2},{p_3},{p_4})\delta(p_1+p_2+p_3+p_4) \\= &\frac{1}{2} \Big( g_{adbc}\big({p_1}, {p_4}| {p_2}, {p_3} \big) + g_{bcad}\big({p_2}, {p_3} | {p_1}, {p_4} \big) - g_{bdac}\big({p_2}, {p_4} | {p_1}, {p_3} \big)- g_{acbd}\big({p_1}, {p_3} | {p_2}, {p_4} \big)\Big) \delta(p_1+p_2+p_3+p_4)   \, .
\end{split}
\end{align}
The dualised, momentum-dependent Riemann tensor obeys the same (anti-)symmetry properties and Bianchi identities as the familiar Riemann tensor. Any sum over internal field-space indices, \ie contraction, is mapped to $\sum_a \rightarrow \sum_a \int \d p$, where an additional integral over momentum has to be performed. Fields with lowered field-space indices are given by $\phi_a=g_{ab} \phi^b \rightarrow \phi_a(-p)$, so raised and lowered indices correspond to incoming and outgoing momenta. We now want to use this duality to incorporate the momentum-dependent counter-term stemming from renormalisation into the geometric picture and consider a momentum and scale-dependent running of the (kinematic) curvature of the field-space manifold.

\section{Loop Corrections}
\label{sec_oneLoop}

In this Section we will investigate the 1-loop structure of the NLSM in Riemann normal coordinates. The loop corrections of the NLSM in combination with the geometry-kinematics duality will allow us to argue for a universal structure of loop corrections for arbitrary theories of massless bosons. Our starting point is the Lagrangian of the NLSM in RNC as discussed in Section \ref{sec_primer}. For the readers' convenience, we have restated it here:  

\begin{align}
\begin{split}
\label{eq_LaNLSM}
	\cL = \frac{1}{2} \Bigg(\delta_{ab} &+ \frac{1}{3} {R^\text{col}}_{acdb}\, \phi^c\phi^d \Bigg) \partial \phi^a \partial \phi^b \, .
\end{split}
\end{align} 

As we are only interested in the quantum corrections to the $4$-point amplitude, we have neglected the higher-dimensional interaction terms. ${R^\text{col}}_{acdb}$ is the leading-order 4-point coupling without any loop corrections evaluated at the vacuum configuration. We stress again that this interaction is independent of kinematics and so far purely a curvature tensor of the field-space manifold. With this we can compute the $1$-loop $s$-channel amplitude and find

\begin{align}
    \cA^{4}_{\text{1-Loop}} = 2 i\int \frac{d^Dq}{(2 \pi )^D} \Bigg( \frac{  \left( p_3 \cdot q+p_4 \cdot q + q^2-s  \right)^2 }{ 9 \,
   q^{2}
   (p_1+p_2+q)^{2}}\,{R^\text{col}}_{a} {}^{\{mn\}}{}_{b}
  {R^\text{col}}_{c\{mn\}d} \nonumber \\  - \frac{\, (p_1 \cdot q-p_2 \cdot q )(p_3 \cdot q-p_4 \cdot q ) }{  q^{2}
    (p_1+p_2+q)^{2}} \,&{R^\text{col}}_a{}^{[mn]}{}_b
{R^\text{col}}_{c[mn]d}  \Bigg)\, ,
    \label{1loopNLSM1}
\end{align}\\   

which matches the cross-check via on-shell cut-constructed amplitude in \cref{eq_1LoopConstR}. 
Here $\{nm\}$ is the symmetrisation and $[nm]$ the anti-symmetrisation of the colour-indices $(n,m)$, \ie 

\begin{align}
{R^\text{col}}_{a\{mn\}b}=\frac{1}{2}\left({R^\text{col}}_{amnb} +{R^\text{col}}_{anmb}   \right)\, , \qquad  \qquad
{R^\text{col}}_{a[mn]b}=\frac{1}{2}\left({R^\text{col}}_{amnb} -{R^\text{col}}_{anmb}   \right) \, .
\label{rimsim}
\end{align}

We can find the $t$- and $u$- channel $1$-loop contributions via crossing symmetry. For completeness, we have listed them in \cref{eq_1LoopConstR}. From the amplitude in \cref{1loopNLSM1}, one can use the geometry-kinematics duality and obtain any one loop  amplitude of massless bosons. We stress that the duality must be applied at this level, \textit{before} the loop integration\footnote{Also the integral has to be dropped as it will arise from the duality-replacement of the geometric quantities after which $\sum_a \rightarrow \sum_a \int \d p$ as ${R^\text{colour}}_{....} \rightarrow R(p,\ldots)_{....}\;\delta(\bullet)$, one of the momentum $\delta$-functions can be used to solve one of the two resulting integrals and we return to ordinary loop integrals after the geometry-kinematics duality replacement.}$^,$\footnote{In the NLSM, all contractions between symmetrised and anti-symmetrised tensors vanish. Once we invoke the geometry–kinematics duality, this cancellation no longer holds, so we must keep contractions involving both symmetrised and antisymmetrised Riemann tensors before applying the duality.}.

At this point we can make an interesting observation, as the geometry-kinematics duality is defined on the level of the action via \cref{eq_actionDuality}, and holds for loop amplitudes at the \emph{integrand level}, we can see from the action that the Feynman rules will have to be linear combinations of Riemann tensors with momentum dependent prefactors. In \cref{ssec_contractions} we argue that any arbitrary linear combination of index permutations of Riemann tensors can be reduced to the form

\begin{align}
     \sum_{\sigma(a,b,n,m)} a_\sigma R_{\sigma(a,b,n,m)} =  \cA_s R_{a\{nm\}b} + \cA_a R_{a[nm]b} \, ,
\end{align}

therefore the off-shell Feynman rules for a 4-point interaction will take the form

\begin{equation}
    \cF_{abcd} \sim \cA_s(p_i\cdot p_j) R_{a\{cd\}b} + \cA_a(p_i\cdot p_j) R_{a[cd]b} \, ,
\end{equation}

where $\cA_{a,s}$ are functions of the (off-shell) momenta involved in the $4$-point correlation function, and in general can be even or odd under exchange of the momenta as they have to be exchanged in conjunction with the colour indices. Applying usual Feynman rules to derive loop amplitudes, the only possible contraction structures that can show up at $1$-loop level are


\begin{align}
\label{eq_doubleContrs1}
\begin{split}
    \cA^\text{1-Loop}_{abcd}    {\sim \int_k  \Big(\cA_s(p_i\cdot p_j) R_{a\{nm\}b} + \cA_a(p_i\cdot p_j) R_{a[nm]b}  \Big) \Big(\cB_s(p_i\cdot p_j) R_{c\{nm\}d} + \cB_a(p_i\cdot p_j) R_{c[nm]d}  \Big) \, ,}
\end{split}
\end{align}

where $(a,b,c,d)$ are colour-tangent space indices. After applying the geometry-kinematics duality, the dualised Riemann tensor depends on momenta, but it still obeys the same (anti-)symmetry relations and Bianchi identities as the familiar Riemann tensor. As we have only used these properties to argue for the structure in \cref{eq_doubleContrs1}, this structure must persist even after applying the geometry-kinematics duality and before integration. Therefore the form of the 1-loop correction to the curvature tensor is given by a simple contraction structure of dualised Riemann tensors and therefore valid for arbitrary theories of massless bosons.

Similar arguments apply to the structure of two-loop corrections. In \cref{app_2Loop}, we provide symmetry-based reasoning that determines their contraction structure. They are to be thought of as claims made on the \emph{integrand} level of the NLSM. An example of a two-loop integral of the NLSM is given in \cref{eq_tooLazy7}, the other integrals and dualised form are listed in \cref{app_2Loop}. 

\begin{equation}
\label{eq_tooLazy7} \vcenter{\hbox{\includegraphics[width=.18\textwidth]{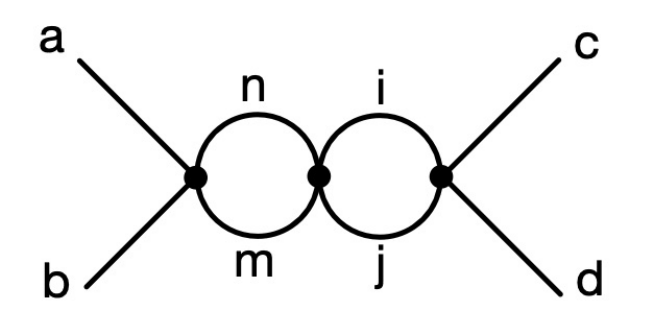}}}\sim \int_{k,q}  \; \big( \cA R_{a\{ij\}b} + \cB R_{a[ij]b}\big)\big( \cC R_{n\{ij\}m} + \cD R_{n[ij]m}\big)\big( \cE R_{c\{nm\}d} + \cF R_{c[nm]d}\big) \, .
\end{equation}

We leave the closer study of the two loop-level for future work.

\subsection{Example: Nambu-Goldstone Theory}

In this Subsection we present how the 1-loop result of the NLSM Lagragian in \cref{1loopNLSM1} together with the geometry-kinematics duality can be used to compute the 1-loop correction of the Nambu-Goldstone theory, which is given by the Lagrangian 

\begin{equation}
    \cL= \frac{1}{2}\Bigg(\delta_{ab}+ \frac{1}{4}\tilde{g}_{abcd}\partial\phi_c\partial\phi_d\Bigg)\partial\phi_a\partial\phi_b  \, .
    \label{NBg1}
\end{equation}

From this, we can read off the dualised metric derivatives

\begin{equation}
    g_{abcd}(p_1,p_2|p_3,p_4) = \frac{1}{2}\,{(p_3 \cdot p_4)} \; \tilde{g}_{abcd} \, , 
    \label{NBgm1}
\end{equation}

where $\tilde{g}_{abcd}$ encodes the colour structure of the interaction and is independent of kinematics.  Using the geometry-kinematics duality, the dualised curvature tensor \cref{eq_genKinCurvTensor}  is given by 

\begin{equation}
    R_{abcd}(p_1,p_2|p_3,p_4) = \frac{1}{4}\,\Big({(p_1 \cdot p_4+p_2 \cdot p_3)} \, \tilde{g}_{adbc} -   {(p_1 \cdot p_3+p_2 \cdot p_4)} \, \tilde{g}_{acbd} \Big) \, . 
    \label{RNBgm1}
\end{equation}

We can now substitute the dualised curvature tensor into our one-loop result for the NLSM, given in \cref{1loopNLSM1}. Again, we have to be careful with raised and lowered indices, as well as the $\sum \rightarrow\sum\int$ replacements and keep the contractions of symmetrised and anti-symmetrised tensors that vanish in the NLSM. 
After making these substitutions and performing the Passarino-Veltman reduction, we obtain

\begin{align}
    \cA^{4}_{\text{1-Loop}} =i\Bigg(  
    -\frac{ D s^4 \left( \tilde{g}_{a\{mn\}b} \tilde{g}_{cdmn} +\tilde{g}_{c\{mn\}d} \tilde{g}_{abmn} \right)}{64 (D-1)}  
    -\frac{  s^3 (t-u) \left( \tilde{g}_{a[mn]b} \tilde{g}_{c[mn]d}  \right)}{32 (D-1)} -\frac{ s^4 \left(\tilde{g}_{abmn} \tilde{g}_{cdmn}  \right)}{32}& \nonumber \\   -\frac{ \Big(\big(D (D + 2) + 2\big) s^4 + 2 (t^2 - u^2)^2\Big)\tilde{g}_{a\{mn\}b} \tilde{g}_{c\{mn\}d} }{128 (D^2-1)} 
&\Bigg)  \cB_0\big(\sqrt{s}\big) \, , 
    \label{1loopNBG1}
\end{align}

where $\{nm\}$ and $[nm]$ refer to the symmetrisation and anti-symmetrisation in the colour indices, as defined in \cref{rimsim} and $\mathcal{B}_0 (p) \equiv \int \frac{d^4k}{(2\pi)^4} \; \frac{1}{k^2(k-p)^2} $ is the scalar bubble integral.

We have explicitly double-checked the result in \cref{1loopNBG1} via direct Feynman diagram computations starting with the Lagrangian in \cref{NBg1}. The direct computations have been performed using FeynRules~\cite{feynrules} routines to generate the Feynman rules for different theories. Then using FeynArts~\cite{feynarts}, we converted them into model files and using FeynCalc~\cite{feync}, we computed amplitudes and reduced 1-loop integrals to Passarino-Veltman integrals, which is just the bubble integral, since we only have $4$-point interactions in the theory. 

We have verified the validity of the duality replacement at the loop level for various Lagrangians, including theories with higher-derivative interactions and those with scalar potentials, and they all agree with the results obtained from standard perturbation theory.

\section{Curvature at 1-Loop Order}
\label{sec_curv}

In this Section, we study how the Riemann tensor and curvature of the field-space manifold change at one loop due to the renormalisation procedure. Firstly, to regularise the amplitude in \cref{1loopNLSM1} we introduce a higher-dimensional operator in the Lagrangian 

\begin{equation}
\label{eq_NLSMwCT}
    \cL = \frac{1}{2}\Bigg(\delta_{ab}+ \frac{1}{3} {R^\text{col}}_{acdb} \;\phi^c \phi^d + \frac{1}{4}\hat{g}_{abcd}\;\partial\phi^c\partial\phi^d\Bigg) \partial \phi^{a} \partial \phi^{b} \, .
\end{equation}

The addition of this higher-dimensional operator, that will eventually absorb the divergence of the $4$-point function, changes the on-shell $4$-point amplitude.  At the leading order the amplitude now reads

\begin{align}
    \cA_{\text{LO}} = s \, {R^\text{col}}_{acbd}+t \, {R^\text{col}}_{abcd}+ \frac{1}{4} \left(s^2
 \hat{g}_{a  b c  d}+t^2 \hat{g}_{a c b d}+u^2 \hat{g}_{a d b c}\right) \, ,
\end{align}

and at next-to leading order it is given by $\cA^{4}_{\text{NLO}} =\cA^{4}_{\text{LO}}+\cA^{4}_{\text{1-Loop}}+\ldots$, where again we will just consider the $s$-channel contribution, as the $t$ and $u$-channel can easily be obtained via crossing symmetry. To renormalise the amplitude, we follow the familiar renormalisation procedure and redefine the tensor $\hat{g}\rightarrow Z\hat{g}=\hat{g}+\delta \hat{g}$ and find the following structure 

\begin{align}
\delta \hat{g}_{abcd}= \frac{1}{24 \pi^2 \varepsilon}\Big({R^\text{col}}_{a [mn] d} {R^\text{col}}_{b [mn] c} + 
 {R^\text{col}}_{a [mn] c} {R^\text{col}}_{b [mn] d} - 
 3 {R^\text{col}}_{a \{mn\} b} {R^\text{col}}_{c \{mn\} d}\Big) \, .
 \label{conterm1}
\end{align}

The counter-term $\delta \hat{g}_{abcd}$ depends on the metric through the definition of the Riemann tensor by \cref{rimt}, \ie just the colour-curvature of the NLSM model. To study how this term introduces scale dependence to the curvature of the NLSM, we compute the renormalisation group equation. For the tensor $\hat{g}_{abcd}$ the Callan-Symanzik equations~\cite{cohen2020scalesseparatedlectureseffective} read:

\begin{equation}
\label{eq_CallanSym}
\mu \frac{ d\hat{g}_{abcd}}{d\mu} =
 2\, \varepsilon \, \hat{g}_{abcd} \, g_{ijkl} \frac{\partial Z_{abcd}}{\partial g_{ijkl}} \, ,
\end{equation}

where the indices $(a,b,c,d)$ are fixed, while the colour indices $(i,j,k,l)$ have to be summed over. This set of equations can be solved for generic $g_{abcd}$, here for the sake of simplicity we can consider

\begin{equation}
    g_{abcd} = \alpha \, \delta _{ab} \delta _{cd}+ \beta \, \delta _{ad} \delta _{bc}+ \beta \, \delta
   _{ac} \delta _{bd} \, ,
\end{equation}

such that the Callan-Symanzik \cref{eq_CallanSym} reduces to 

\begin{equation}
    \mu \frac{ d\hat{g}_{abcd}}{d\mu} = -\frac{\lambda^2 }{6
   \pi ^2} \Big(2 \, \delta _{ad} \delta _{bc}+2 \, \delta
   _{ac} \delta _{bd}+(3 N-7) \, \delta _{ab} \delta _{cd}\Big) \, .
   \label{diffeq1}
\end{equation}

where $N$ is the number of colours and thus the dimension of the field-space manifold, therefore $\delta _{ab} \delta ^{ab} = N$ and we defined $\beta - \alpha  = \lambda$. It is straightforward to solve the differential \cref{diffeq1} and we find for the running coupling

\begin{equation}
   \hat{g}_{abcd}(\mu) =\hat{g}^0_{abcd} - \frac{ \lambda^2}{6
   \pi ^2}\,\text{log}(\mu) \Big(2 \, \delta _{ad} \delta _{bc}+2 \, \delta
   _{ac} \delta _{bd}+(3 N-7) \, \delta _{ab} \delta _{cd}\Big) \, .
   \label{NLSM1sol1}
\end{equation}

Here we defined $\hat{g}^0_{abcd}$ as the initial value for $\hat{g}_{abcd}$ at $\mu=1$. The counter term in the Lagrangian given in \cref{eq_NLSMwCT}  runs with the renormalisation scale $\mu$, and we would like to express this in the geometric picture. Since the counter-term carries additional derivatives, it will have additional momentum dependence. To give an interpretation to the geometrised counter-term we lean on the geometry-kinematics duality defined on the level of the action as given in \cref{eq_actionDuality}. 
With this, the RG-improved Riemann tensor is given by

\begin{equation}
  \bar{R}_{abcd}(\mu)= {R^{\text{col}}}_{abcd}+   \frac{1}{4}(p_{a}\cdot p_{d}+p_{b}\cdot p_{c})
   \hat{g}_{adbc}(\mu)- \frac{1}{4}(p_{a}\cdot p_{c}+p_{b}\cdot p_{d})
   \hat{g}_{acbd}(\mu) \, .
\end{equation}

Assuming the initial condition
\begin{align}
    \hat{g}^0_{abcd}=  \frac{\lambda^2}{6 \pi^2 }  \left(\delta _{ad} \delta _{bc}+\delta _{ac} \delta_{bd}+\delta _{ab} \delta _{cd} \right)
\end{align}
for the coupling in \cref{NLSM1sol1}, we can explicitly write $\bar{R}_{abcd}(\mu)$ for the NLSM as 

\begin{align}
\label{eq_NLSMCurvRunning}
\begin{split}
\bar{R}_{abcd}(\mu)  =& -\lambda  \left(\delta _{ad} \delta _{bc}-\delta _{ac} \delta
   _{bd}\right)  \\& +  \frac{\lambda ^2 }{6 \pi
   ^2}\left(p_a\cdot p_d+p_b\cdot p_c\right)
   \bigg(\delta_{ad} \delta_{bc}+\delta
   _{ac} \delta_{bd}+\delta_{ab} \delta_{cd} -\log (\mu) \Big(2 \delta_{ac} \delta
   _{bd}+2 \delta_{ab} \delta_{cd}+(3 N-7) \delta_{ad} \delta
   _{bc}\Big)\bigg)  \\
 &  -\frac{\lambda ^2 }{6 \pi ^2}\left(p_a\cdot p_c+p_b\cdot
   p_d\right)  \bigg(\delta_{ac} \delta
   _{bd}+\delta_{ad} \delta_{bc}+\delta_{ab} \delta
   _{cd}-\log (\mu) \Big(2
   \delta_{ad} \delta_{bc}+2 \delta_{ab} \delta_{cd}+(3 N-7)
   \delta_{ac} \delta_{bd}\Big)\bigg) \, .
\end{split}
\end{align}

 and putting the momenta on shell, the Ricci tensor and scalar are given by 

\begin{align}
\begin{split}
R_{ab}(\mu) & =\lambda  (N-1)\delta _{ab}\Bigg(1+\frac{\lambda  }{6 \pi ^2}\bigg(\frac{(N+2)}{(N-1)}
   (t-u)  - \log (\mu)
    \Big((3 N-4) t-5 u\Big)\bigg)\Bigg) \, , \\
R(\mu) & = \underbrace{\lambda  N(N-1)}_{S^N\text{ curvature}} \Bigg( 1 +\frac{\lambda  }{6 \pi ^2}N \bigg(\underbrace{\frac{(N+2)}{(N-1)} (t-u)}_{\text{momentum dependence}}- \underbrace{\log (\mu) \Big((3 N-4) t-5 u\Big)}_{\text{running curvature}}\bigg)\Bigg) \, .
\end{split}
\end{align}

\begin{figure}
    \centering
    \begin{minipage}{0.45\textwidth}
        \centering
        \begin{tikzpicture}[scale=0.4]
            \def\R{4} 
            \def\angEl{25} 
            \def\angPhiOne{230} 
            \pgfmathsetmacro\H{\R*cos(\angEl)} 
            \LongitudePlane[pzplane]{\angEl}{\angPhiOne}
            \fill[ball color=white!10] (0,0) circle (\R); 
            \coordinate[mark coordinate] (O) at (0,0);
            \coordinate[mark coordinate] (N) at (0,\H);
            \coordinate[mark coordinate] (S) at (0,-\H);
            \path[pzplane] (0:\R) coordinate (P);
            \DrawLongitudeCircle[\R]{\angPhiOne}
            \DrawLatitudeCircle[\R]{0} 
            \draw[-,dashed, thick] (N) -- (S);
            \draw[->,blue,thick] (O) -- node[above]{$R$} (P);
            \LatitudePlane[equator]{\angEl}{0}
            \draw[equator,thick] (-180:\R) to[bend right=45] (-90:\R);
            \draw[equator,thick] (-90:\R) to[bend right=45] (0:\R);
            \LongitudePlane[pzplane]{\angEl}{\angPhiOne}
            \node at (0, \R+1) {\Large \textbf{tree-level}};\textbf{}
        \end{tikzpicture}
    \end{minipage}
    \begin{minipage}{0.45\textwidth}
        \centering
        \begin{tikzpicture}[scale=0.4]
            \def\R{6} 
            \def\angEl{25} 
            \def\angPhiOne{230} 
            \pgfmathsetmacro\H{\R*cos(\angEl)} 
            \LongitudePlane[pzplane]{\angEl}{\angPhiOne}
            \fill[ball color=white!10] (0,0) circle (\R); 
            \coordinate[mark coordinate] (O) at (0,0);
            \coordinate[mark coordinate] (N) at (0,\H);
            \coordinate[mark coordinate] (S) at (0,-\H);
            \path[pzplane] (0:\R) coordinate (P);
            \DrawLongitudeCircle[\R]{\angPhiOne}
            \DrawLatitudeCircle[\R]{0} 
            \draw[-,dashed, thick] (N) -- (S);
            \draw[->,blue,thick] (O) -- node[above right, pos=0.6, xshift=-11pt, yshift=10pt]{$R(s,t;\mu)$} (P);
            \LatitudePlane[equator]{\angEl}{0}
            \draw[equator,thick] (-180:\R) to[bend right=45] (-90:\R);
            \draw[equator,thick] (-90:\R) to[bend right=45] (0:\R);
            \LongitudePlane[pzplane]{\angEl}{\angPhiOne}
            \node at (0, \R+1) {\Large \textbf{one-loop}};\textbf{}
        \end{tikzpicture}
    \end{minipage}
    \caption{Schematic representation illustrating the variation of the curvature of the field-space manifold as a function of $s,t$ and $\mu$ when considering loop contributions. The radius $R(s,t;\mu)$ ( $\sim$(Ricci scalar)$^{-\nicefrac{1}{2}}$) encapsulates how the geometric structure of the field-space manifold evolves with these parameters, emphasizing the dependence of curvature on the renormalisation scale and interaction kinematics.}
    \label{fig_curvatureFct}
\end{figure}
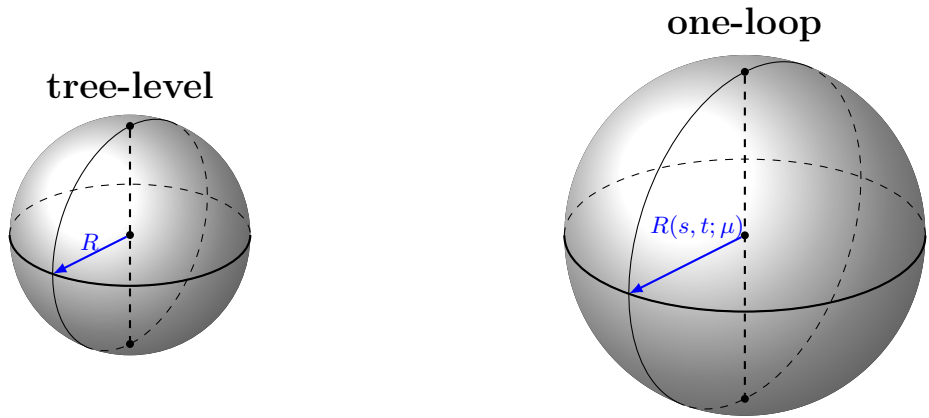

We see that the geometric invariants of the field-space manifold at one-loop level acquire a dependence on the renormalisation scale $\mu$ and the kinematics of the scattering. While the coupling now runs, the curvature tensor in \cref{eq_NLSMCurvRunning} still obeys the form $R \sim t^\rho - u^\rho$ as required by the (anti-)symmetry properties and the Bianchi identity\footnote{After setting the momenta on-shell.}. The Ricci scalar of $S^N$ is given by $R\big(S^N\big) = \frac{N(N-1)}{r^2}$, therefore we are describing a sphere with radius $r=\lambda^{-\nicefrac{1}{2}}$, this is not surprising as the field-space manifold for the pion Lagrangian is given by a sphere of dimension $N$. As the field-space manifold is the $S^n$, we can schematically visualise the running of the kinematic curvature and the change of the field-space manifold in Figure \ref{fig_curvatureFct}.

\subsection{Example: Coloured \texorpdfstring{$\phi^4$}{TEXT}-Theory}

Following the same procedure as outlined above for the pion Lagrangian, we can compute the one-loop Riemann tensor for other theories as well. Here we present a coloured $\phi^4$ interaction, where $\cL_\text{int} \supset \frac{1}{4!} \;g_{abcd}\;(\phi^4)^{abcd}$. For this theory the RGE for the coupling is given by

\begin{equation}
    \mu \frac{ dg_{abcd}}{d\mu} = \frac{ g_{adnm} g_{bcnm}+g_{acnm}
   g_{bdnm}+g_{abnm} g_{cdnm}}{32 \pi ^2} \, .
   \label{rgescalq}
\end{equation}
\Cref{rgescalq} can be solved for different choices of the tensor $g$, for sake of simplicity we consider the case
\begin{equation}
   g_{abcd} = \frac{\lambda(\mu) }{3} \, (\delta _{ad} \delta _{bc}+\delta _{ac} \delta _{bd}+\delta _{ab}
   \delta _{cd}) \, ,
\end{equation}

where $g_{abcd} = \lambda $ corresponds to the single colour scenario. Using this choice for the tensor $g$, the renormalisation group equations simplify to

\begin{equation}
   \mu \frac{d\lambda}{d\mu} = \frac{\lambda(\mu)^2 }{32 \pi^2}\frac{(N+8)}{3}   \, ,
\end{equation}

and by assuming $\lambda(1)=\lambda$, we find the running of $\lambda$ given by

\begin{equation}
\lambda(\mu) = \frac{\lambda }{1-\frac{\lambda   }{32 \pi ^2}\frac{(N+8)}{3} \log (\mu )} \, ,
\end{equation}

So far we have only followed the ordinary renormalisation procedure of a coloured $\phi^4$-theory. Now we can apply the geometry-kinematics duality, reading off the dualised metric derivatives by comparing to \cref{eq_actionDuality} and then computing the Riemann tensor via \cref{eq_genKinCurvTensor}. With this we find that the RG-improved Riemann tensor now reads

\begin{align}
\bar{R}_{abcd}(t,u;\mu)&=\frac{\lambda(\mu)  }{36
  } \left(\delta _{ad} \delta _{bc}+\delta _{ac}
   \delta _{bd}+\delta _{ab} \delta _{cd}\right)
   \left(\frac{1}{t}-\frac{1}{u}\right) \, , 
\end{align}

and the running Ricci scalar is given by

\begin{equation}
  \bar{R}(t,u;\mu) =   \lambda(\mu)\, \frac{N(N+2)}{9}\,\Bigg(\frac{1}{ t }-\frac{1}{ u } \Bigg) \, .
\end{equation}

As for the NLSM model discussed above, we again find the geometric invariants to acquire a dependence on the renormalisation scale $\mu$ in addition to the kinematics. While the coupling now runs, the on-shell curvature tensors still obey the form $R \sim t^\rho - u^\rho$ as is required by the (anti-)symmetry properties and the Bianchi identity. This result is a nice cross-check, as $\phi^4$ is renormaliseable we do not need to introduce a higher dimensional local counter-term with momentum dependence and $g_{abcd}$ has to acquire a scale dependence. As expected, in the $N=1$ scenario, the Riemann tensor extends the form found in the single-colour scenario discussed in \cite{Cheung:2022vnd} to the loop level.

\section{Discussion}

In the geometric picture of scalar effective field theories, particularly in the nonlinear sigma model, the scalar fields are interpreted as coordinates on a curved target space, described by a Riemannian field-space manifold. The dynamics of the theory are governed by a field-dependent metric on this manifold, which determines the form of the Lagrangian and encodes the curvature. Therefore, the geometric structure determines interactions through curvature-dependent terms involving the Riemann tensor and its covariant derivatives. This perspective provides a natural way to understand the structure of interactions and the organization of the EFT expansion in a covariant manner, where physical observables are determined via the geometric invariants of the field-space manifold.

In this work, we have explored the geometric formulation of scalar effective field theories, emphasizing how quantum corrections at the loop level modify the underlying field-space geometry. Adopting the nonlinear sigma model as our starting point, we computed the one-loop amplitude in the geometric language, uncovering a universal contraction structure of curvature tensors at 1- and 2- loop level that encode the quantum corrections of the theory. This approach highlights how quantum corrections affect the field-space geometry of the NLSM.

Furthermore, by leveraging the recently uncovered geometry-kinematics duality, we argued that the structure of these quantum corrections is not unique to the NLSM but extends more generally to arbitrary theories of massless bosons. These include theories with scalar potentials, higher-derivative theories, as well as theories involving vector and tensor fields. Our findings suggest that the renormalisation of such theories can be systematically understood through geometric invariants, with the curvature of the field-space manifold playing a central role in determining interaction corrections. After computing the loop corrections and introducing appropriate counter-terms to renormalise the theory, we examined the scale dependence of the geometric invariants under the renormalisation group flow. This analysis provides insight into how the field-space metric and its associated curvature evolve at different energy scales.

Since the geometry-kinematics duality applies not only to scalars but also to vectors and tensors, our results should naturally extend beyond the realm of scalar field theories. In particular, this geometric perspective can be applied to (non-) abelian gauge theories and theories of gravity. A detailed exploration of these directions, including the interplay between renormalisation group flows and the curvature of the gauge and gravitational field-space manifolds, is an exciting avenue we leave for future studies.

\section*{Acknowledgments}

We would like to thank Tim Cohen and Andreas Helset for very illuminating discussions about the geometry-kinematics duality and we are grateful for their comments on the draft. Also we are very grateful to Lihang Zhou for helpful suggestions and pointing out typos. This work has been supported by the Collaborative Research Center SFB1258, the Munich Institute for Astro-, Particle and BioPhysics (MIAPbP), and by the Excellence Cluster ORIGINS, which is funded by the Deutsche Forschungsgemeinschaft (DFG, German Research Foundation) under Germany’s Excellence Strategy – EXC 2094 – 39078331. The work of LB is supported in part by the U.S.
Department of Energy under Grant No. DE- SC0010102 and by the College of Arts and Sciences of Florida State University. LB work was performed in part at the Aspen Center for Physics, which is supported by a grant from the Simons Foundation (1161654, Troyer).
The research of DH is partially supported by the International Max Planck
Research School (IMPRS) on “Elementary Particle Physics”.
LB thanks the Technische Universität München (TUM) for the hospitality and  the IMPRS for partial support during the completion of this work.

\numberwithin{equation}{section}
\begin{appendices}

\section{On-Shell Methods}
\label{app_OnShell}

Amplitude methods have recently been used to systematically study low-energy effective field theories (EFTs), and unitarity methods have been applied to explore the renormalisation group (RG) structure of EFTs~\cite{Bern:1994zx,Caron-Huot:2016cwu,Craig:2019wmo,Bern:2019wie,EliasMiro:2020tdv,Jiang:2020mhe,Bern:2020ikv,Cheung:2015aba,Jiang:2020rwz,Baratella:2021guc,Aebischer:2025zxg,Bresciani:2024shu,Bresciani:2023jsu}. In this Appendix, we review these methods and demonstrate how they can be applied to extract the bubble coefficient.

The analytic structure of scattering amplitudes consists of poles, polynomials (from tree-level diagrams), and logarithms (from loops). We can probe this structure employing unitarity of the $S$-matrix and the (generalised) optical theorem, which gives the relation for the transfer matrix $\text{Disc} \:T = 2 \: \text{Im}\: T = T^\dagger T$. Where imaginary part should be interpreted as a discontinuity across a branch cut singularity of the amplitude. Considering this in a perturbative expansion in a coupling constant we can collect the terms of matching powers in $\lambda$ and find $\disc \: T_4^{(0)} = 0$, and $\disc \: T_4^{(1)} = {T_4^{(0)}}^{ \dagger} T_4^{(0)}$. We see that tree amplitudes do not exhibit any branch-cuts and that the imaginary part of the 1-loop amplitude arises from its discontinuities across branch-cuts and is given as a product of tree amplitudes, as a result of the (generalised) optical theorem. The right hand side includes an implicit sum over all possible intermediate states (species, helicity, flavour, ...) which may lie between the $T$ matrices as well as a phase-space integration. In analogy to the treatment of tree-level amplitudes in the on-shell paradigm, unitarity provides us with an interpretation of factorisation of the loop amplitude into lower-loops/trees when the loop momenta go on-shell. Generalised unitarity extends this by allowing complex loop momenta, leading to unitarity cuts that decompose loop amplitudes into products of tree-level amplitudes via Cutkosky rules~\cite{Bern:2011qt}. The generalisation to complex momenta will allow us to put a higher number of propagators on-shell simultaneously, yielding singularities located away from the physical region~\cite{nima_simplestQFT}. This extends the familiar Cutkosky rules and unitarity cuts to complex momenta, where propagators are cut by performing the replacement $(p^2 - m^2 +\im \epsilon)^{-1} \rightarrow - 2\pi \, \delta^{(+)}(p^2-m^2)$, which gives the double cut of the amplitude as

\begin{equation*}
    \text{Cut}_2\:  \cA^{\text{1-loop}} = \int d \Pi \; \cA_{L}^{\text{tree}} \times \cA_{R}^{\text{tree}}.
\end{equation*}

Now in oder to study the amplitude via unitarity cuts, one has to rewrite the one-loop amplitude according to the Passarino-Veltman (PV) decomposition~\cite{PVdecomp}, which provides a complete set of basis integrals for 1-loop scattering amplitudes. According to this decomposition any 1-loop amplitude in $D$ spacetime dimensions can be expressed as a sum of $m$-gon scalar loop integrals $I_m$ with $m= 1,2,...,D$ as

\begin{equation*}
\label{eq_PVReduction}
    \cA^{1-loop} = \sum_i B_D^i \, I_D^i + \sum_j B_{D-1}^j I_{D-1}^j + \ldots + \sum_k B_2^k \, I_2^k + \text{rational parts}\, .
\end{equation*}

\begin{figure}[ht]
    \centering
    \includegraphics[width=0.5\textwidth]{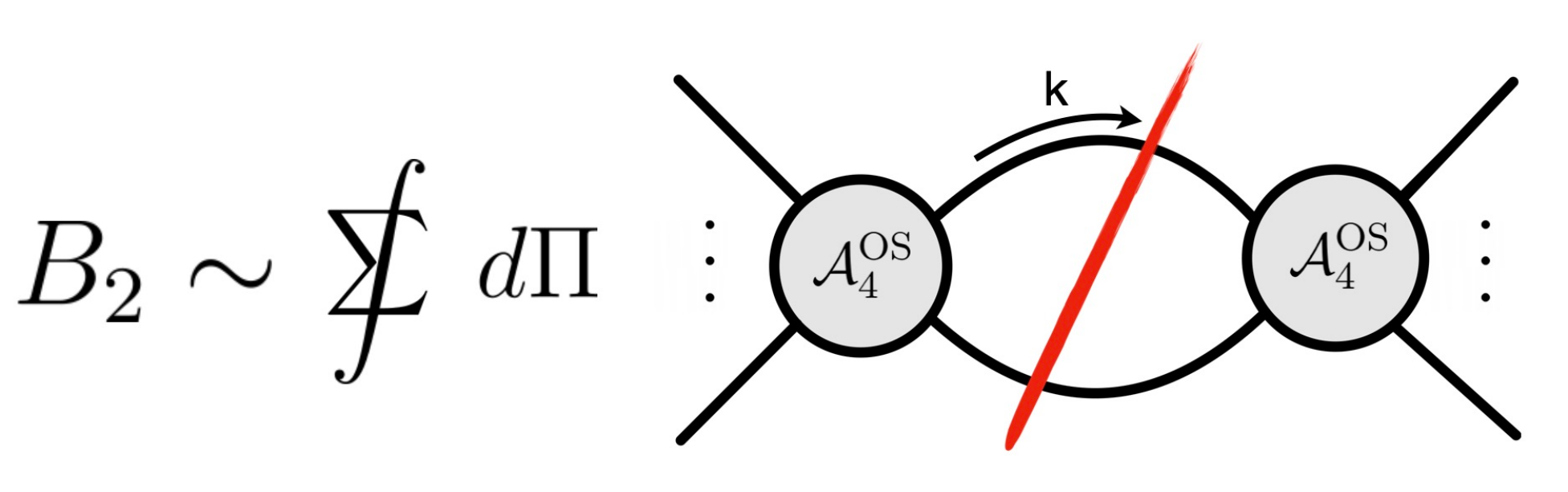} 
    \caption{In theories with only 4-point interactions the bubble coefficient can be extracted from the double cut of the loop amplitudes directly.}
    \label{fig_bubbleCut}
\end{figure}

We do not need to compute the decomposition explicitly, the knowledge of its existence suffices as our goal will be to extract the anomalous dimension via cuts. The coefficients $C_m^i$ are rational functions of the external kinematical invariants, the colour and flavour structure of the amplitude under consideration. As rational functions, they do not exhibit branch cuts and are unaffected by cutting of the amplitude. They encode all the kinematic information of the external states in the expansion into the $m$-gon scalar 1-loop integrals $I_m^i$. The scalar loop integrals are defined in dimensional regularisation while their coefficients and the rational parts are kept in $D$ integer dimensions~\cite{nima_simplestQFT}. The unitarity method bypasses explicit Feynman loop calculations by comparing the analytical structures of both sides of \cref{eq_PVReduction}. Using generalized unitarity, the branch-cut structure of the loop amplitude is matched with that of scalar master integrals, allowing the extraction of the desired expansion coefficients $B^i$. In the 4-dimensional case the box and triangle integrals are UV finite and only the bubble integral carries a UV divergence and thus any contribution to the anomalous dimension must come from the bubble coefficients. Thus it suffices to extract the bubble coefficients $B_2$ to study anomalous dimensions. The idea is to compare the $2$-cut of the amplitude to the $2$-cut of the Passarino-Veltman expansion via

\begin{equation}
\label{eq_cutsInPV}
     \left[ \text{Cut}^{(i)}_2 \right] \cA = \sum_j B^j_d \left[ \text{Cut}^{(i)}_2 \right] I ^j_d + \ldots +  \sum_i B^i_2 \left[ \text{Cut}^{(i)}_2 \right] I ^i_2,
\end{equation}

where the LHS is determined through unitarity via an appropriate product of tree amplitudes, schematically shown in Figure \ref{fig_bubbleCut}, and the cuts of the $m$-gon scalar integrals on the RHS are known~\cite{britto_review}. \Cref{eq_cutsInPV} is the key to the generalized unitarity method, as cuts of the full loop amplitude are related to tree amplitudes, while on the other side of the equation, only a subset of the master integrals allows cuts for a given kinematical configuration. In the case of the NLSM theory the RHS is particularly simple as we only have 4-point amplitudes and thus only bubble integrals. Therefore, we get the relation for the bubble coefficient directly without worrying about interference from triangle and box coefficients. This is schematically shown in Figure \ref{fig_bubbleCut}.

Now we shall get an expression for the anomalous dimension of amplitudes from the bubble coefficient. An amplitude at 1-loop order can be written as $\cA = \cA_{\mathcal{O}_i} + \cA_{\text{1-loop}}$. Here $\cA_{\mathcal{O}_i}$ is a purely local amplitude expressed as a polynomial in the kinematic invariants. Given the amplitude in this form and our knowledge that we can decompose $\cA_{\text{1-loop}}$ according to Passarino-Veltman, we gain the anomalous dimension by demanding that the full amplitude is independent of the renormalisation scale, which is introduced by the bubble integral. The only other part of the amplitude that depends on the scale is the Wilson coefficient $\wilsonCoeff{i}$ of the local amplitude, this requirement is stated as

\begin{equation*}
    0 = \frac{d \cA}{d \log \mu} = \frac{d}{d \log \mu}\cA_{\mathcal{O}_i}  + \frac{d}{d \log \mu} \cA_{\text{1-loop}} = \gamma_i \frac{\cA_{\mathcal{O}_i}}{\wilsonCoeff{i}} + \sum_j B_2^{(j)} \frac{d I_2^{(2)}}{d \log \mu}\, .
\end{equation*}

We know the behaviour of the scalar bubble integral to be

\begin{equation*}
\label{eq_bubbleInt1}
    I_2^{(j)} = \frac{1}{16 \pi^2} \left( \frac{1}{\epsilon} + \log \left( \frac{\mu^2}{-P_j^2} \right) + \orderOf{\epsilon} \right),
\end{equation*}

and thus we see that the anomalous dimension is related to the bubble coefficient via

\begin{equation*}
    \gamma_i \cA_{\mathcal{O}_i} = -\frac{1}{8\pi^2} \sum_j B_2^{(j)} + \gamma_{\text{IR}} \cA_{\mathcal{O}_i}.
\end{equation*}

The IR divergences are absent in scalar theories with 4-point interactions~\cite{Catani_1996,Catani_1997,Catani_1998,Becher_2009,Becher_2009_1}, which is exactly the class of theories that we are considering, so the bubble coefficient fully determines the UV divergence and we do not have to take IR divergences into account when computing it with on-shell methods. Therefore we can employ generalised unitarity methods and perform a double-cut of the amplitude in order to extract the bubble coefficient via 

\begin{equation*}
\begin{split}
     B_2^{(i)} &= \sum_j B_2^{(j)}\underbrace{\left[ \text{Cut}_2^{(i)}\right] I_2^{(j)}}_{\delta_{ij}} \\ &=\left[ \text{Cut}_2^{(i)}\right] \cA = \sumint \; d \Pi \; \cA_L^{\text{OS}}(\ldots, \ell_1, \ell_2) \times\cA_R^{\text{OS}} ( - \bar{\ell_2}, - \bar{\ell_1}, \ldots)  \, .
\end{split}
\end{equation*}

Here we see that unitarity demands that the sum of residues of all cuts in the PV expansion must be equivalent to the product of 2 on-shell amplitudes~\cite{elvang_book}. In addition, we have already used that the double-cut of the scalar bubble $\text{Cut}_2 \, I_2 = 1$, this is as expected a purely rational constant. Here, $\cA_{L/R}$ describe tree-level on-shell amplitudes. We have to sum over all possible cuts, meaning all distinct distributions of the external states $1,\ldots,n$ onto the subamplitudes $\cA_{L/R}$. We also have to sum over all possible internal states joining the two on-shell amplitudes, where we have chosen to employ crossing symmetry on the states in $\cA_R$ to abide with our all-incoming convention. This leads to the negative momenta in $\cA_R$ and the notation $\bar{\ell}_i$ which means that all quantum numbers have to be inverted with respect to those of the state labeled by $\ell_i$. In order to deal with the spinors of negative momenta we employ the convention $|-\lambda \ra = i|\lambda \ra$ and $|-\lambda ] = i|\lambda]$. The phase space integral is normalised as $\int d\Pi =  1$.

With all this information, we can determine the bubble coefficient of the s-channel 1-loop contribution by finding the on-shell amplitudes at either side of the cut as 

\begin{align*}
	\cA_L^{\text{OS}} &= \cA_4^{\text{OS}} \big(p_1^a, p_2^b, -k^n, (k-p_{12})^m\big) = -2 \, {R^\text{col}}_{abnm} \, {p_1}\cdot{k} +  {R^\text{col}}_{anbm} \, s\, , \\
	\cA_R^{\text{OS}} &= \cA_4^{\text{OS}} \big(k^n, (-k-p_{34})^m, p_3^c,p_4^d\big) =2 \, {\big({R^\text{col}}\big)^{nm}}_{cd}\, \, {p_3}\cdot{k} +  {{{\big({R^\text{col}}\big)^n}_c}^m}_{d} \, s\, , 
\end{align*}

and then computing the phase-space integral over all possible internal states, where we employ a convenient parametrisation of the internal momentum and the phase-space~\cite{Caron-Huot:2016cwu} 

\begin{align*}
    k \ra = \cos \theta \: 1 \ra &- \sin \theta \: e^{\im \phi} \: 2\ra \qquad \& \qquad k ] = \cos \theta \: 1] - \sin \theta \: e^{-\im \phi} \: 2] \, ,  \\
	&\int d\Pi = \int_0^{2\pi} \, \frac{d\phi}{2\pi} \, \int_0^{\nicefrac{\pi}{2}} \, d\theta \; 2 \, \sin\theta \, \cos\theta \, , 
\end{align*}

and thus find

\begin{align}
\label{eq_bubbleInt}
\begin{split}
	B_2^{(s)} &= \frac{i}{2}\int d\Pi \, \cA_L^{\text{OS}} \times  \cA_R^{\text{OS}} \\
    &=\frac{i}{2}\; s\,\int d\Pi \, \Big( \big( {R}_{anbm} - \sin^2\theta \,{R}_{abnm} \big) \, \big((t \cos^2 \theta + u \sin^2 \theta) \,{{R}^{nm}}_{cd} + s\,{{{{R}^n}_c}^m}_{d} \big)\Big) \, \\
    &=\frac{i}{2}\; s\int d\Pi\; s\, {R}_{a\{nm\}b}{{{R}_{c}}^{\{nm\}}}_{d} + (s+2\,t)\cos^2{2\theta}\;{R}_{a[nm]b}{{{R}_{c}}^{[nm]}}_{d} 
\end{split}
\end{align}

where repeated indices on the Riemann tensors have to be summed over to account for all possible intermediate-state particles. Employing generalised unitarity we see that the bubble coefficients of all amplitudes can be written as:

\begin{align}
    B_2^{(s)} =\int d\Pi \; \Big( f_1(s,t;k)\, R_{a\{nm\}b}
   {{R_{c}}^{\{nm\}}}_{d}+f_2(s,t;k)\, R_{a[nm]b}
   {{R_{c}}^{[nm]}}_{d}\Big) \, ,
    \label{1loopgenApp}
\end{align}

where $f_1$ and $f_2$ are functions of the Mandelstam variables that depend on the kinematic part of the metric. This structure is expected as these are all independent two-contractions of two Riemann tensors, \cf discussion in Appendix \ref{ssec_contractions}. Since we have only employed symmetries of the Riemann tensor to arrive at this form, this should also hold for kinematics dependent Riemann tensors, as they satisfy the same symmetry properties Riemann tensors on Riemannian manifolds. For the kinematics independent curvature in the NLSM model, as in \cref{eq_LaNLSMinRNC1}, we find

\begin{equation*}
	B_2^{(s)} =    \frac{i}{2}\Big(s^2  \, {R}_{a\{nm\}b}
   {{{R}_{c}}^{\{nm\}}}_{d}+\frac{1}{3} s\, (\,s+2\, t)\,{R}_{a[nm]b}
   {{{R}_{c}}^{[nm]}}_{d}\Big)\, .
\end{equation*}

With this result for the $s$-channel bubble coefficient, we can employ crossing symmetry to find the other bubble coefficients. Then the 1-loop amplitude for kinematics-independent Riemann Tensor of the NLSM is given by

\begin{align}
\begin{split}
\label{eq_1LoopConstR}
	\mathcal{A}_{\text{1-loop}}( p_1^a,p_2^b,p_3^c,p_4^d) = 
		&\frac{i}{2}\Big({R}_{a\{nm\}b}
   {{{R}_{c}}^{\{nm\}}}_{d} \,s ^2+ \frac{1}{3} \,{R}_{a[nm]b}
   {{{R}_{c}}^{[nm]}}_{d} \;s\,(s+2\,t) \Big) \mathcal{B}_{0} (\sqrt{s}) \\
		&+ \frac{i}{2}\Big({R}_{a\{nm\}c}
   {{{R}_{b}}^{\{nm\}}}_{d} \, t^2 +\frac{1}{3} \,{R}_{a[nm]c}
   {{{R}_{b}}^{[nm]}}_{d} \, t\,  (t+2\, s) \Big) \mathcal{B}_0 (\sqrt{t}) \\
		&+ \frac{i}{2}\Big( {R}_{a\{nm\}d}
   {{{R}_{c}}^{\{nm\}}}_{b} \, u^2 +\frac{1}{3} \, {R}_{a[nm]d}
   {{{R}_{c}}^{[nm]}}_{b} \, u\,  (u+2\, t) \Big) \mathcal{B}_0 (\sqrt{u}) \,.	
\end{split}
\end{align}

Where $\mathcal{B}_0 (p) \equiv \int \frac{d^4k}{(2\pi)^4} \; \frac{1}{k^2(k-p)^2} $ is the scalar bubble integral.

\section{Why Only These Contractions?}
\label{ssec_contractions}

The 24 possible permutations of the indices of \( R_{abnm} \) are:
\begin{align*}
    & R_{abnm}, R_{abmn}, R_{anbm}, R_{anmb}, R_{ambn}, R_{amnb}, \\
    & R_{banm}, R_{bamn}, R_{bnam}, R_{bnma}, R_{bmna}, R_{bman}, \\
    & R_{nabm}, R_{namb}, R_{nbam}, R_{nbma}, R_{nmab}, R_{nmba}, \\
    & R_{mabn}, R_{manb}, R_{mban}, R_{mbna}, R_{mnab}, R_{mnba}.
\end{align*}

\noindent The Riemann tensor \( R_{abnm} \) satisfies the following symmetries:

\begin{align*}
    R_{abnm} &= -R_{banm} \quad \text{(Antisymmetry in first two indices)} \\
    R_{abnm} &= -R_{abmn} \quad \text{(Antisymmetry in last two indices)} \\
    R_{abnm} &= R_{nmab} \quad \text{(Symmetry under exchange of index pairs)} \\
    0&= R_{abnm} + R_{anmb} + R_{ambn} \quad \text{(First Bianchi identity)}
\end{align*}

\noindent Taking the symmetries and anti-symmetries into account, a linear combination of possible permutations reduces to:

\begin{align*}
     \sum_{\sigma(a,b,n,m)} a_\sigma R_{\sigma(a,b,n,m)} = \cA R_{abnm}+ \cB R_{anbm}+\cC R_{ambn}\, .
\end{align*}

\noindent Where the terms on the RHS can further be related via the first Bianchi identity. Thus we can express the RHS in terms of $R_{anmb}$ and $R_{amnb}$ using the Bianchi identity. 

\begin{align*}
	\cA R_{abnm}+\cB R_{anmb} + \cC R_{amnb} = (\cB - \cA) R_{anmb} + (\cC+\cA) R_{amnb}
\end{align*}

\noindent Decomposing the tensors into symmetric and anti-symmetric parts we can write this as

\begin{align*}
	(\cB - \cA) R_{a\{nm\}b} +(\cB - \cA) R_{a[nm]b} + (\cC+\cA) R_{a(mn)b}+ (\cC+\cA) R_{a[mn]b} \\
	= \underbrace{(\cB + \cC)}_{\equiv \cA_s} R_{a\{nm\}b} + \underbrace{(\cB - \cC-2\cA)}_{\equiv \cA_a} R_{a[nm]b} \, .
\end{align*}

\noindent Therefore, an arbitrary linear combination of all possible permutations of indices can always be reduced to

\begin{align}
\label{eq_permSum}
     \sum_{\sigma(a,b,n,m)} a_\sigma R_{\sigma(a,b,n,m)} =  \cA_s R_{a\{nm\}b} + \cA_a R_{a[nm]b} \, ,
\end{align}

\noindent and therefore all possible double contractions\footnote{To simplify notation we do not distinguish between co- and contravariant indices.} of arbitrary linear combinations of Riemann tensors are given by

\begin{align}
\label{eq_doubleContrs}
\begin{split}
    \Bigg( \sum_{\sigma(a,b,n,m)} a_\sigma R_{\sigma(a,b,n,m)} \Bigg) \Bigg( \sum_{\sigma(c,d,n,m)} b_\sigma R_{\sigma(c,d,n,m)} \Bigg)
     &= \big( \cA_s R_{a\{nm\}b} + \cA_a R_{a[nm]b}  \big)\big( \cB_s R_{c\{nm\}d} + \cB_a R_{c[nm]d}\big) \\
\end{split}
\end{align}

\section{2-Loop Structure}
\label{app_2Loop}

We know from arguments above that any possible linear combination of permutations of Riemann tensor indices can be written as in \cref{eq_permSum}. This is very useful as the off-shell Feynman rules for the $4$-point vertex are given precisely by linear combinations of permutations of Riemann tensors as in equation \eqref{eq_permSum} with momentum product dependent functions $a_\sigma (p_i \cdot p_j)$, $\cA (p_i \cdot p_j)$ and $ \cB(p_i \cdot p_j)$. Thinking about the possible contraction structure of 3 Riemann tensors with 4 open indices we find that they correspond to the Feynman diagrams that we can draw at 2-loop order in the NLSM. Here we list the contractions of the Riemann tensors that correspond to loop corrections for the two-loop $s$-channel diagrams.

\begin{equation}
\label{eq_tooLazy3}
    \vcenter{\hbox{\includegraphics[width=.15\textwidth]{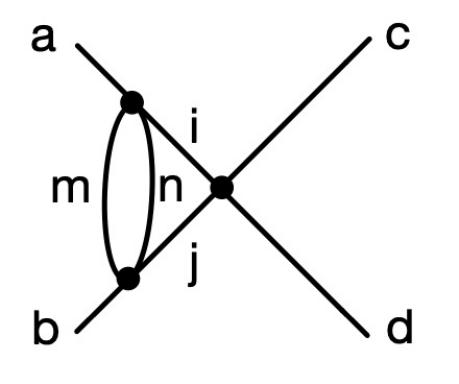}}}\sim \big( \cA R_{a\{nm\}i} R_{b\{nm\}_j}+ \cB R_{a[nm]i}R_{b[nm]j}\big)\big( \cC R_{c\{ij\}d} + \cD R_{c[ij]d}\big)\, .
\end{equation}

\begin{equation}
\label{eq_tooLazy4}
    \vcenter{\hbox{\includegraphics[width=.15\textwidth]{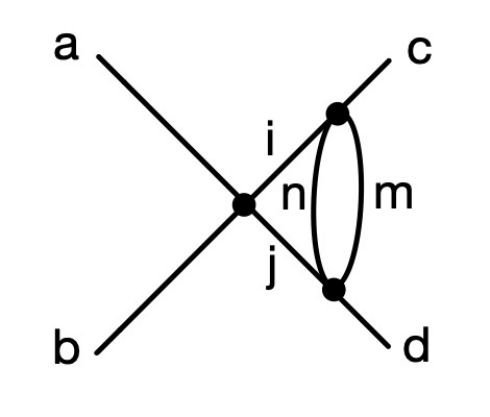}}} \sim \big( \cA R_{a\{ij\}b} + \cB R_{a[ij]b} \big) \big( \cC R_{c\{nm\}i} R_{d\{nm\}j }+ \cD R_{c[nm]i}R_{d[nm]j} \big)   \, .
\end{equation}

\begin{equation}
\label{eq_tooLazy5}
    \vcenter{\hbox{\includegraphics[width=.18\textwidth]{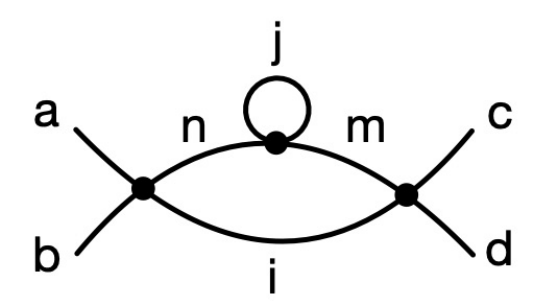}}} \sim \big( \cA R_{a\{ni\}b} + \cB R_{a[ni]b} \big) R^{nm} \big( \cC R_{c\{mi\}d}+ \cD R_{c[mi]d} \big) \, .
\end{equation}

\begin{equation}
\label{eq_tooLazy6}
    \vcenter{\hbox{\includegraphics[width=.18\textwidth]{doubleCandy.pdf}}}\sim \big( \cA R_{a\{nm\}b}R_{c\{ij\}d} R_{n\{ij\}m} + \cB R_{a[nm]b} R_{c[ij]d} R_{n[ij]m} \big) \, .
\end{equation}

We observe that even though the structure increases in complexity compared to the $1$-loop case, we can still reduce the contractions of the Riemann tensors to a quite simple expression. Here $\cA(p_i \cdot p_j), \cB(p_i \cdot p_j), \cC(p_i \cdot p_j), \cD(p_i \cdot p_j)$ are scalar functions of momentum products (including loop momenta) resulting from the off-shell Feynman rules and are \ig different for each diagram\footnote{For notational simplicity we used $\cA, \cB, \cC, \cD$ for every diagram, but they are \ig inequivalent for the different diagrams.}. Particularly simple structure of the diagram in \cref{eq_tooLazy6} can be seen as (anti-) symmetrisation of the inner indices also implies (anti-) symmetrisation in the outer indices\footnote{$R_{a(nm)b} = \nicefrac{1}{2}(R_{anmb}+R_{amnb}) = \nicefrac{1}{2}(R_{mban}+R_{nbam}) = \nicefrac{1}{2}(R_{bmna}+R_{bnma}) =  R_{b(nm)a}  $}$^,$\footnote{$R_{c[nm]b} = \nicefrac{1}{2}(R_{cnmb}-R_{cmnb}) = \nicefrac{1}{2}(R_{mbcn}-R_{nbcm}) = \nicefrac{1}{2}(R_{bmnc}-R_{bnmc}) =  -R_{b[nm]c}  $}. We recover the $t$- and $u$-channel expressions via appropriate permutation of the external colour indices $(a,b,c,d)$. It is important to highlight that here we make the claim about this structure at the integrand level of the NLSM. The other possible contractions in equations \eqref{eq_tooLazy1} and \eqref{eq_tooLazy2} give vacuum bubbles and corrections to the external legs respectively, these diagrams/contraction structures are not important when considering loop corrections to the $4$-point amplitude.

\begin{equation}
\label{eq_tooLazy1}
    \vcenter{\hbox{\includegraphics[width=.1\textwidth]{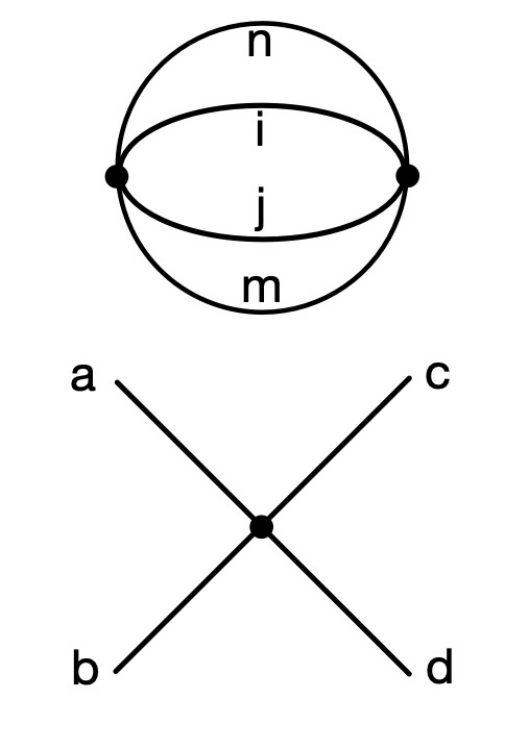}}} \sim R_{abcd} R_{nmij} R_{nmij}\, .
\end{equation}

\begin{equation}
\label{eq_tooLazy2}
    \vcenter{\hbox{\includegraphics[width=.1\textwidth]{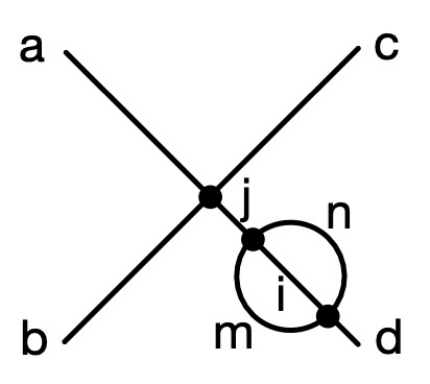}}} \sim R_{abcj}{R}_{jnmi}{R_{nmid}}\, .
\end{equation}

If we apply the geometry-kinematics duality we expect the general form

\begin{equation}
    \cA \sim \int_{k,q} \big( \cA R_{a\{ij\}b} + \cB R_{a[ij]b}\big)\big( \cC R_{n\{ij\}m} + \cD R_{n[ij]m}\big)\big( \cE R_{c\{nm\}d} + \cF R_{c[nm]d}\big) 
\end{equation}

\end{appendices}

\bibliographystyle{JHEP.bst}
\bibliography{geo_kin}

\end{document}